\documentclass[%
 reprint,
 amsmath,amssymb,
aps,
pra,
]{revtex4-1}

\usepackage{graphicx}
\usepackage{dcolumn}
\usepackage{bm}

\begin{document}

\preprint{APS/123-QED}
\title{Stepwise relaxation and stochastic precession in degenerate oscillators dispersively coupled to particles }
\author{Christin Rh\'en}
\email{christin.rhen@chalmers.se}
\author{Andreas Isacsson}

\affiliation{Department of Physics\\
Chalmers University of Technology\\
SE-412 96 G\"oteborg\\
Sweden}

\date{\today}

\begin{abstract}
By numerical integration, we study the relaxation dynamics of degenerate harmonic oscillator modes dispersively coupled to particle positions. Depending on whether the effective inertial potential induced by the oscillators keep the particles confined, or if the particle trajectories traverse the system, the local oscillator energy dissipation rate changes drastically. The inertial trapping, release and retrapping of particles results in a characteristic step-wise relaxation process, with alternating regions of fast and slow dissipation. To demonstrate this phenomenon we consider first a one-dimensional minimal prototype model which displays these characteristics. We then treat the effect of dispersive interaction in a model corresponding to an adsorbate diffusing on a circular membrane interacting with its three lowest vibrational modes. In the latter model, stepwise relaxation appears only in the presence of thermal noise, which also causes a slow-in-time stochastic precession of the mixing angle between the degenerate eigenmodes.
\end{abstract}

\pacs{Valid PACS appear here}
\maketitle

\section{Introduction}
Physical systems displaying discrete resonances are conveniently modelled by a set of harmonic oscillators. Interactions with a surrounding medium, e.g. a thermal environment, is usually modelled through coupling of the oscillator degrees of freedom to some set of bath coordinates. Common and well-studied examples are resonant systems coupled to two-level systems~\cite{Maizelis_2014} or to a quasi-continuous bath of thermal harmonic oscillators~\cite{DykmanKrivoglaz, Weissbook}. Coupling the system to a thermal bath allows it to relax to the thermal equilibrium state while also, in accordance with the fluctuation-dissipation theorem, inducing thermal noise. 

The coupling between a system and its environment can also have an indirect component, proceeding via an auxiliary system. The dissipative dynamics can then change dramatically. This was recently demonstrated experimentally for micro- and nanoelectromechanical systems (MEMS/NEMS), where intermediate coupling via internal resonances caused marked qualitative changes in the ring-down dynamics~\cite{Chen_2016,Guttinger_2016,Shoshani_2017}. Here, we present a general theoretical model for harmonic oscillators indirectly coupled to the environment, through an auxiliary system in the shape of a set of free particle coordinates.

The model in question consists, in its most general form, of an assembly of $N$ harmonic oscillators $q_n(t)$, with natural frequencies $\Omega_n$. The auxiliary system is comprised of $K$ ``particles'' with positions $x_k(t)$:
\begin{equation}
H_0=\frac{1}{2}\sum_{n=1}^N\left(\dot{q}_n^2+\Omega_n^2q_n^2\right)+\frac{\mu}{2}\sum_{k=1}^K \dot{x}_k^2.\label{eq:h0}
\end{equation}
Coupling to the environment is achieved through adding thermal noise and dissipation to the particle equations of motion. We assume the coupling between the particles and the oscillator modes to be dispersive, and of the type
\begin{equation}
H_{\rm int}=-\frac{1}{2}\sum_{k,n,m} g_{mn}{q}_n{q}_m\phi_{n}(x_k)\phi_{m}(x_k).\label{eq:hi}
\end{equation}
Here, $g_{mn}=g_{nm}$ are coupling constants, and $\phi_n(x)$ functions that depend on the particle coordinate. Eqs.~\eqref{eq:h0}-\eqref{eq:hi} are reminiscent of the type of Hamiltonian usually studied in the context of dissipation in open systems. It should be noted, though, that in that context, the oscillators are considered to represent the degrees of freedom of the thermal environment, leading to relaxation and thermalization of the system in coordinate space $x_k$. In contrast, we are here concerned with the relaxation dynamics of a small number of discrete oscillator modes $q_n$, as they interact via and with the particle coordinates. 


Physically, this class of Hamiltonians \eqref{eq:h0}-\eqref{eq:hi} finds application in, for instance, the context of NEMS, where the oscillator degrees of freedom $q_n$ correspond to vibrational eigenmodes and $x_k$ denote the positions of adsorbed particles diffusing on the resonator. The $\phi_n$ are then associated with the spatial mode functions. The frequency noise associated with the fluctuating mass distribution and the resulting back-action on the particles from the resonator motion have been shown to change both the resonant response and the characteristics of the dissipative dynamics~\cite{Dykman_2010,Atalaya_2011,Atalaya_2011_2,Yang_2011,Atalaya_2012,Barnard_2012,Edblom_2014,Jiang_2014,Schneider_2014,Rhen_2016}. While this type of system is of great interest for applications such as ultrasensitive mass-sensing~\cite{Lassange_2008,Jensen_2008,Chaste_2012}, it is also generic enough to find applications in other fields~\cite{Rhen_2017}, and is in its own right worth studying.


Although several studies of NEMS-particle systems have been done, the effect of mode degeneracy has not yet been considered. This open question is becoming increasingly relevant as the state of the art develops to the point where such modes can be resolved~\cite{Tsioutsios_2016}. 

Here, we find that even a minimal model, consisting of two degenerate harmonic oscillator modes coupled to a single, weakly damped particle that induces a mode coupling, exhibits highly nontrivial relaxation dynamics. The total oscillator energy decays in a stepwise manner, with plateaus of very low dissipation appearing at similar total energies, regardless of initial conditions. In these regions, the two modes perform coherent coupled oscillations, while the particle is trapped at an antinode of the oscillator motion. In between the plateaus, the total oscillator energy, as well as the energy of each mode, decays rapidly, and the particle traverses its entire domain in an erratic manner. We also show how a two-dimensional (2D) equivalent model, inspired by particles adsorbed on circular drum resonators, exhibits analogous relaxation dynamics: stepwise dissipation associated with trapping and retrapping of the particles. Here, however, it is necessary to add thermal fluctuations to the particle dynamics in order for the trapping-retrapping to occur. In that case, and with only a higher degenerate mode excited, the trapped particle incurs a slow stochastic precession of the mixing angle. We characterize the low-frequency noise of this precession.

\section{Stepwise relaxation in a minimal model\label{sec:gen}}
In this section, we restrict the general equations of motion to the, in this context, smallest possible number of degrees of freedom. 
Namely, we set $N=2$ and $K=1$, and consider degenerate modes: $\Omega_1=\Omega_2\equiv\Omega$.  
The coupling matrix is set to $g_{mn}(x)=g$, and we use the mode functions $\phi_1(x)=\sqrt2\cos\pi x$, $\phi_2(x)=\sqrt{2}\sin2\pi x$. The system is normalized to unit length, $|x_k|\le 1/2$, with reflecting boundary conditions. In this model, the effect we seek to illustrate is present also when $T=0$. Hence, we omit thermal noise. 

The equations of motion for this minimal model are
\begin{align}
    \ddot q_1+\left[\Omega^2-g\phi_1^2\right]q_1-g\phi_1\phi_2q_2&=0,\label{eq:q1}\\
    \ddot q_2+\left[\Omega^2-g\phi_2^2\right]q_2-g\phi_1\phi_2q_1&=0,\label{eq:q2}\\
    \ddot x+\Gamma\dot x-\frac g{2\mu}\partial_x\left[\left(q_1\phi_1+q_2\phi_2\right)^2\right]&=0\label{eq:x}.
\end{align}
The structure of these equations is deceptively clear; each oscillator experiences a frequency shift that depends on the particle's position, and an $x$-dependent linear mode coupling appears. The particle is subject to a potential $U(x,t)\propto (q_1\phi_1+q_2\phi_2)^2$. This force is akin to the inertial force a mechanical resonator exert on an adsorbed particle. Consequently, we will refer to it as inertial force. Despite the apparent clarity, as we here demonstrate, the nonlinearity of Eqs.~\eqref{eq:q1}-\eqref{eq:x} means that this simple model exhibits rich and nontrivial relaxation dynamics.

This richness partly arises due to the interference in the potential term $U(x,t)$, which depends not only on the magnitude of the amplitudes $q_{1,2}(t)$, but also their relative phase. For degenerate oscillators, this phase changes slowly in time with a rate depending on particle position. Hence, where a single mode-system (or a non-degenerate one) may have stable trapping points for the particles, points corresponding to the antinode(s) of the mode function, mode degeneracy leads to a radically different situation, with release and retrapping dynamics. 

An analytical approach to the dynamics of the system is a daunting task.
Instead, we have performed extensive numerical simulations by integrating the equations of motion~\eqref{eq:q1}-\eqref{eq:x}, while tracing the time evolution of mode energies, $E_n(t)=\dot q_n^2(t)+\Omega_n^2q_n^2(t)$, $n=1,2$, as well as the total oscillator energy $E=E_1+E_2$, during relaxation. 

\begin{figure*}
    \centering
    \includegraphics[width=\textwidth]{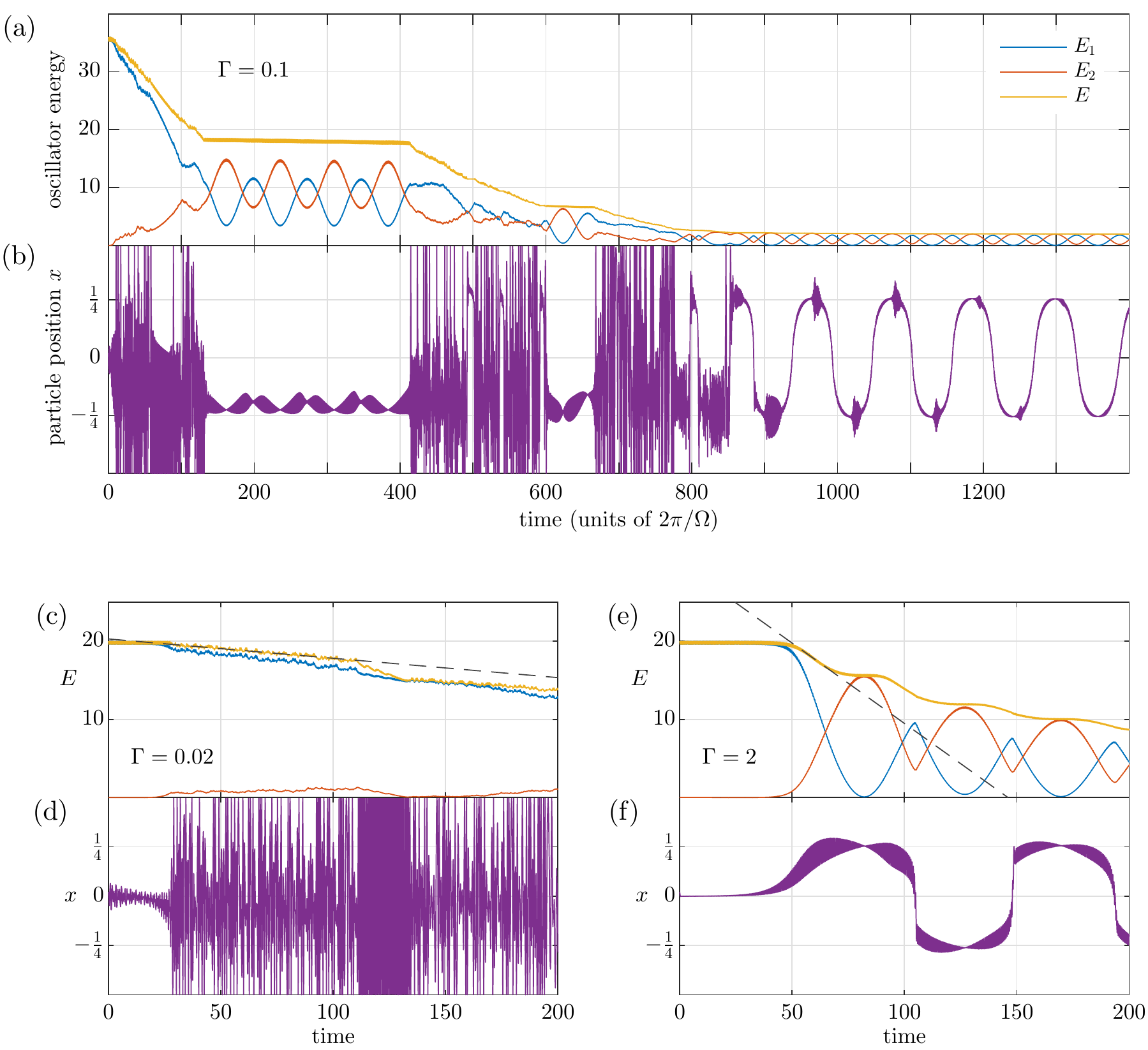}
    \caption{\textbf{(a)} Time evolution of the energies of individual oscillators described by Eqs.\eqref{eq:q1}-\eqref{eq:x}, along with their total energy, at an intermediate decay rate $\Gamma=0.1$. The stepwise decay of total oscillator energy $E$ is clear, as well as the coherent coupled oscillations of the low-dissipation energy plateaus. \textbf{(b)} The corresponding evolution of the particle coordinate $x$. In regions where the particle traverses its entire domain $[-1/2, 1/2]$, the energy $E$ decays rapidly. At the energy plateaus, $x$ is trapped near a minimum of the inertial potential created by the oscillators. For low enough oscillator energy (weak inertial trapping), the particle may switch between minima, but the system dynamics is nonetheless predictable and regular. \textbf{(c)}-\textbf{(d)} Typical system evolution at low decay rate $\Gamma$. Initially, the particle is trapped and the energy decays slowly. After this transient, the particle escapes and moves erratically, corresponding to a rapid decay of energy, indicated by the gray dashed line. The exception to the unpredictable particle motion is the region at approximately $t\times \Omega/2\pi \sim 110 -130 $. Here, the particle motion synchronizes with the oscillators, and ``bounces'' between the boundaries of its domain once every half-period. The dissipation rate of oscillator energy increases correspondingly. \textbf{(e)}-\textbf{(f)} Typical system evolution at high decay rate $\Gamma$. The particle adiabatically follows a minimum of the inertial potential, alternating between the antinodes of $\phi_2(x)$. The corresponding energy relaxation is stepwise but smooth. In each step, the slope is indicated by the gray dashed line.}
    \label{fig:etot}
\end{figure*}

While the details of the dynamics depend on initial conditions, certain features are recurring. For demonstration of these features, the system ring-down for three different values of the damping rate $\Gamma$ is shown in Fig.~\ref{fig:etot} ($g=0.01$, $\mu=1$). The initial condition used is nonzero $q_1(0)$ and $x(0)$, while $q_2(0)=\dot q_2(0)=\dot q_1(0)=\dot x(0)=0$. For intermediate $\Gamma$, [see Fig.~\ref{fig:etot}~(a)-(b)], the most striking feature is the stepwise relaxation that results as the system shifts between regions of very rapid decay of total oscillator energy $E$, and plateaus of very low dissipation. The details of the system behaviour is dramatically different in these two cases. The rapid-decay regions correspond to large particle motion across the entire particle domain, while the near-conservative regions are characterized by particle trapping and coherent, coupled oscillations in the energy of individual oscillators. The extremes of low [Fig.~\ref{fig:etot}~(c)-(d)] and high [Fig.~\ref{fig:etot}~(e)-(f)] damping rates can be roughly understood as sampling only the region of high dissipation or only the plateaus, respectively. 

In the remainder of this Section, we consider some salient features of the minimal model more in detail.

\subsection{Coherent oscillations at plateaus \label{sec:flat}}
In the low-dissipation plateaus, the oscillator relative energy dynamics is reasonably well described by the rotating wave approximation (RWA).  In the RWA, the total energy $E_1+E_2$ remains conserved, but it is nonetheless a convenient framework for studying the coherent oscillations of the individual oscillator energies.

Changing variables [$q_n=a_ne^{i\Omega t}+a_n^*e^{-i\Omega t}$, $\dot q_n=i\Omega(a_ne^{i\Omega t}-a_n^*e^{-i\Omega t})$] and discarding rapidly oscillating terms gives the RWA equations
\begin{align}
    \dot a_1&=-\frac{ig}{2\Omega}\left(\phi_1^2a_1+\phi_1\phi_2a_2\right),\label{eq:a1}\\
    \dot a_2&=-\frac{ig}{2\Omega}\left(\phi_1\phi_2a_1+\phi_2^2a_2\right),\label{eq:a2}
\end{align}
along with analogous equations for the complex conjugates $a_1^*,a_2^*$, and a transformed equation for $x$. The amplitudes $a_{1,2}$ describe the slow evolution of the oscillator amplitudes, and hence vary on the same timescale as the oscillator energies $E_{1,2}$.

Numerically integrating the RWA equations of motion~\eqref{eq:a1}-\eqref{eq:a2} reproduces the coherent oscillations between $E_1$ and $E_2$ seen at the plateaus in Fig.~\ref{fig:etot}~(a). The frequency of these slow oscillations is given by
\begin{equation}
    \lambda\approx \frac{g}{2\Omega}\left(\phi_1^2(x_{\rm eq.})+\phi_2^2(x_{\rm eq.})\right),
\end{equation}
where $x_{\rm eq.}$ is the value near which the particle is trapped. Setting $x_{\rm eq.}\approx \pm0.2$, the period time of coherent oscillations becomes $\lambda^{-1}\approx 64\times 2\pi/\Omega$. This is reasonably close to the slow periods $74\times 2\pi/\Omega$, $68\times 2\pi/\Omega$, and $61\times 2\pi/\Omega$ seen in the three plateaus of Fig.~\ref{fig:etot}~(a). 

\subsection{Robust location of energy plateaus}
\begin{figure}
    \centering
    \includegraphics[width=\columnwidth]{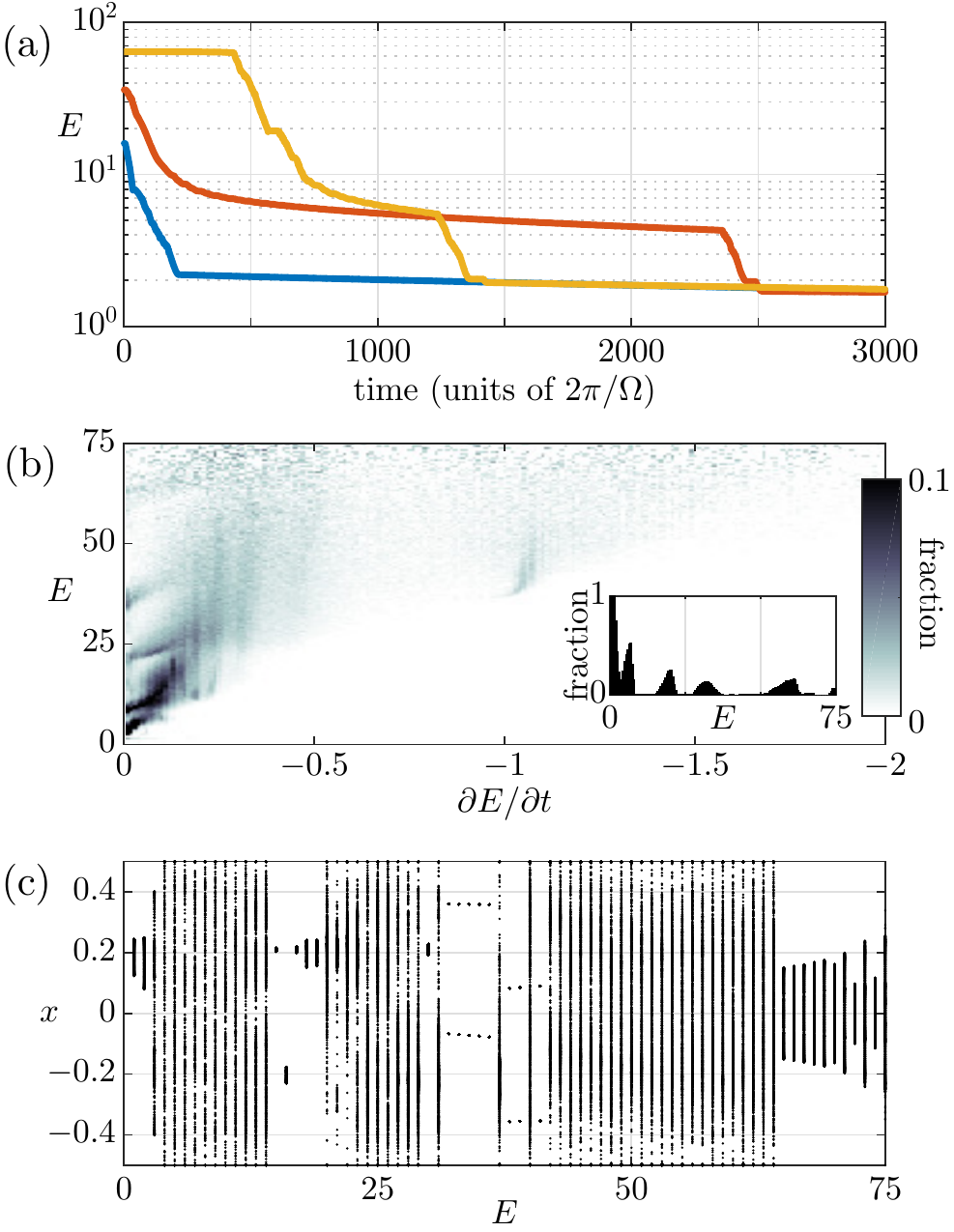}
    \caption{\textbf{(a)} Stepwise decay of total oscillator energy, at different initial conditions and $\Gamma = 0.2$. The low-dissipation plateaus tend to coincide, despite very different starting energies. \textbf{(b)} Distribution of dissipation rates during the ringdown, as a function of energy, compiled from $10^5$ trajectories with random initial conditions. The data for each value of $E$ is normalized to one; the color indicates the fraction of time spent at a certain energy that is also spent at a certain dissipation rate $\partial E/\partial t$. Fractions equal to or greater than $0.1$ are coloured black. Inset: Distribution of energies for which $\partial E/\partial t<0.05$: the low-dissipation plateaus of the ring-down. \textbf{(c)} Periodicity diagram of the particle motion for a wide range of oscillator energies; each point is the position $x(t)$ when $\dot x(t)=0$. Hence, exactly two points for a given $E$ indicates that the particle here moves periodically with the same frequency as the oscillators; the two points are the turning positions of the particle.}
    \label{fig:ehist}
\end{figure}

In the plateau state, when the particle is trapped, energy dissipates slowly until particle release. Quite surprisingly, repeated ring-down simulations at various initial energies reveals that, as shown in Fig.~\ref{fig:ehist}~(a), the energy plateaus seemingly appears at approximately the same total energy $E$ regardless of the initial condition. That is, particle trapping and release appears to predominantly occur around certain total oscillator energies. 

To verify this, $10^5$ trajectories with random initial conditions were simulated, each for $10^7$ oscillator periods. The energy decay rate $\partial E/\partial t$ was calculated at each timestep. For each value of $E$ the distribution of derivatives, across all trajectories, was normalized to 1. The result is shown in Fig.~\ref{fig:ehist}~(b). For low energies there is a clear structure, further illuminated by the inset, which shows the energies at which dissipation rates below $0.05$ (plateaus) occur. While the locations of the plateaus are likely to change with the system parameters (here, $g=0.01$, $\mu=1$, and $\Gamma=0.2$), their locations in energy space do not average out upon random sampling of the initial parameter space. In this sense, they are robust.

The corresponding particle dynamics were examined by numerically solving an energy-conserving version of Eqs.~\eqref{eq:q1}-\eqref{eq:x} for a wide range of initial energies. In practice, at each time step $t_j$ of the integration, the oscillator coordinates $q_n(t_j),\dot q_n(t_j)$ were normalized by a factor $\sqrt{E(0)/E(t_j)}$. In this way, the total oscillator energy of the system is kept constant while the mode coupling and particle dynamics are unchanged. We thus ensure that the system will remain in a certain regime of interest. While the same total-energy conserving effect can be obtained by using the RWA-equations (\ref{eq:a1})-(\ref{eq:a2}), using the full equations, but rescaling energy, allows short-time correlations to be retained.

For each value of the total energy, the periodicity of the particle motion was examined by plotting the $x$-coordinate whenever $\dot x=0$, during the last 300 periods ($2\pi/\Omega$) of a simulation lasting 1500 periods. The result is a diagram that indicates how well the particle motion synchronizes with the oscillators, as well as delineates $\Delta x=\max x-\min x$; the region traversed by the particle. The fact that, at energies where ringdown plateaus appear, the particle is trapped near the antinodes of $\phi_2(x)$ at $\pm 0.2$ is further corroborated. Between these trapped regions there seems to be no pattern to the particle motion, suggestive of chaotic dynamics in the high-dissipation regime. Finally, the particle trapping near $x=0$, the antinode of $\phi_1(x)$, at the right of the diagram reflects that, for these high energies, it takes more time to excite the $q_2$-mode, and the first energy plateau of the system is quasi-stationary during the time-span simulated here.

\begin{figure}
    \centering
    \includegraphics[width=\columnwidth]{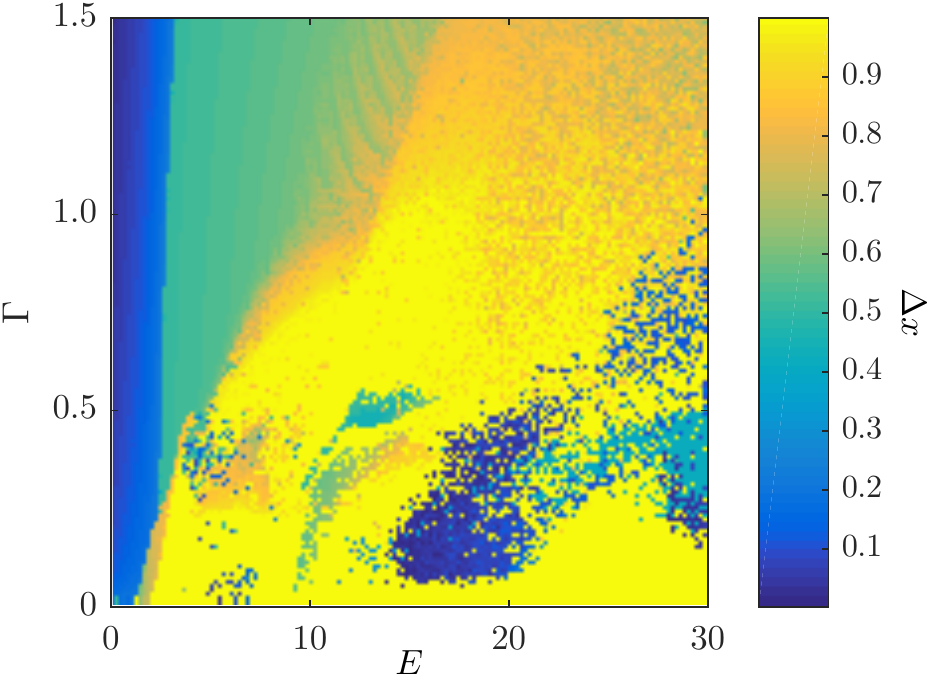}
    \caption{The distance $\Delta x$ between minimum and maximum values of $x$ (compare the particle trajectories in Fig.~\ref{fig:etot}) as a function of damping rate $\Gamma$ and oscillator energy $E$. For a given damping rate, the ringdown of the system follows a horizontal line from right to left in the diagram, crossing trapped, stable regions (blue) and untrapped, erratic ones (yellow). Teal regions, corresponding to $\Delta x\approx 0.5$, result from trajectories where the particle switches
    between the two potential wells at $\pm 0.25$, and thus represent trapped states.    \label{fig:stability}}
\end{figure}


Investigating also the effect of the decay rate $\Gamma$, in Fig.~\ref{fig:stability}, we map the trapped and erratic regimes of the system, as determined by $\Delta x$, as a function of $E(0)$ and $\Gamma$. 
Two primarily trapped regions can be seen (coloured blue in Fig.~\ref{fig:stability}), that are reminiscient of the Arnold tongues of the Mathieu equation. This is not unexpected; systems very similar to Eqs.~\eqref{eq:q1}-\eqref{eq:x} have been shown~\cite{Edblom_2014} to be related to parametric oscillators, who are in turn described by the Mathieu equation. Here, however, the distribution of trapped and untrapped states is fractal-like, recalling chaotic dynamics. 

\subsection{Rapid dissipation of oscillator energy\label{sec:linear}} 

\begin{figure}
    \centering
    \includegraphics[width=\columnwidth]{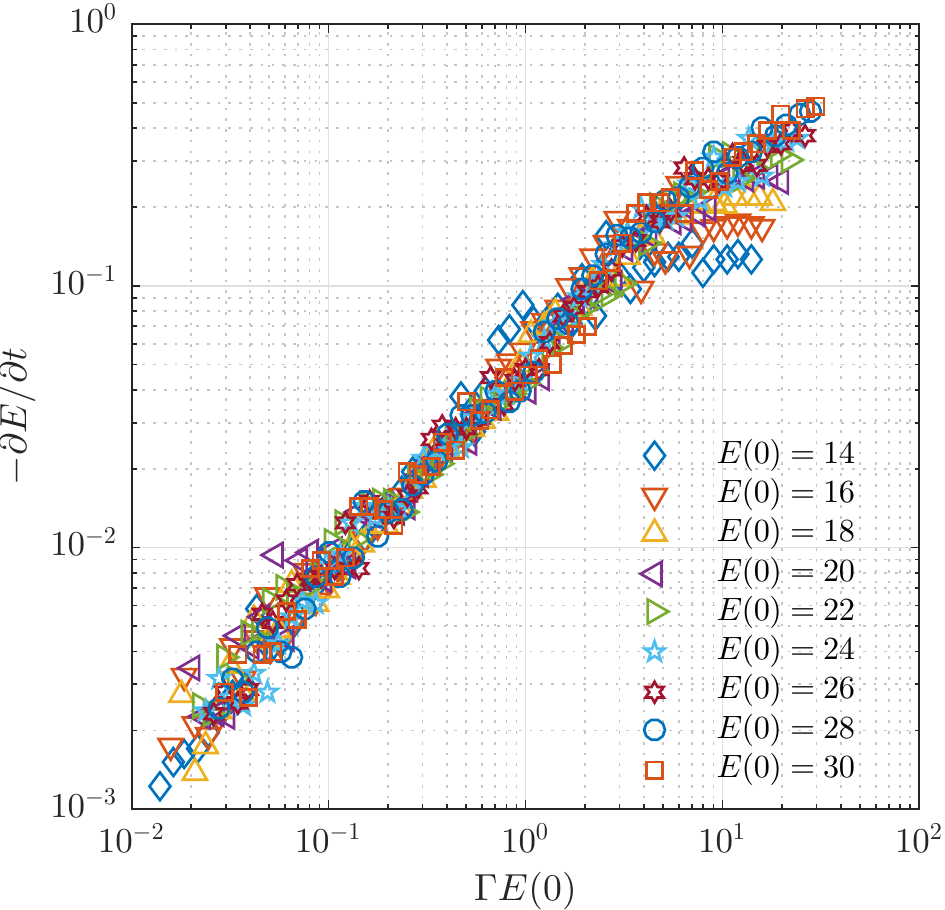}
    \caption{Dissipation $-\partial E/\partial t$ of total energy in the first high-dissipation region, as a function of $\Gamma E(0)$. The dissipation rate was determined by extracting the slope of the approximately linear decay of energy, indicated by grey dashed lines in Fig.~\ref{fig:etot}~(c) and (e).}
    \label{fig:drops}
\end{figure}

We now turn to the dynamics between the plateaus where energy is rapidly dissipated. In this regime, the decay of the oscillator is approximately linear in time for short times (see Fig.~\ref{fig:etot}). 
We investigated the energy decay rate in the high-dissipation region by extracting the initial slope  [the gray dashed lines in Fig.~\ref{fig:etot}~(c) and (e)] for a wide range of initial oscillator energies $E(0)$ and particle damping rates $\Gamma$. As can be seen in Fig~\ref{fig:drops}, the energy decay rate $\partial E/\partial t$ is linear in  $E(0)\Gamma$ for $E(0)\Gamma<1$. For higher values, a clear deviation can be seen.

The only mechanism of energy loss present in the system ~\eqref{eq:q1}-\eqref{eq:x} is the term $\mu\Gamma\dot x$ of Eq.~\eqref{eq:x}, corresponding to a total average energy loss rate proportional to the average particle kinetic energy: $\mu \Gamma \left< {\dot x}^2\right>=2\Gamma E_{\rm p}$. Thus, the scaling $\partial E/\partial t\propto  -E(0)\Gamma$ implies that the average kinetic energy of the particle is proportional to the oscillator energy at the onset of erratic motion, $E_{\rm p}\propto E$, which results in an exponential decay of energy. This behavior stems from the irregularity of the particle motion, which effectively renders the forcing term in \eqref{eq:x} to act as a noise source, causing "thermalization" of the particle motion. Numerically, we find that during the erratic periods, $E_{\rm p}(t)=\mu\left<\dot{x}^2\right>/2=\alpha E(t)/2$, where the proportionality factor is of the order $\alpha \sim g/\mu$. 

This decay mechanism, leading to exponential relaxation, should be contrasted to what we have previously reported~\cite{Edblom_2014,Rhen_2016}; when adsorbed particles diffuse on the surface of a nanomechanical resonator, the resonator vibrational energy decays in a linear manner if the resonator amplitude is high enough. In that case, linear-in-time dissipation coincided with the particles being inertially trapped, and as they escaped to diffuse across their entire domain, the dissipation rate changed to nonlinear. As erratic motion from the thermal stochastic force drove particles from the equilibrium position, work was done to bring them back, causing dissipation of resonator energy. The mechanism is here similar, but because the effective noise is proportional to oscillator energy rather than a fixed temperature, the relaxation becomes exponential instead. 

\section{Relaxation and stochastic precession in a membrane model\label{sec:graphene}}
The model system treated above is inherently one-dimensional (1D) in the coordinate space $x$ of diffusing particles. While 1D realizations where degenerate oscillator modes have different mode functions exist, notably modes tuned to a crossing or narrow anticrossing, they are quite uncommon. However, by extending the situation to higher dimensions, modal degeneracy can appear from symmetry considerations. 

\begin{figure}[t]
    \centering
    \includegraphics[width=\columnwidth]{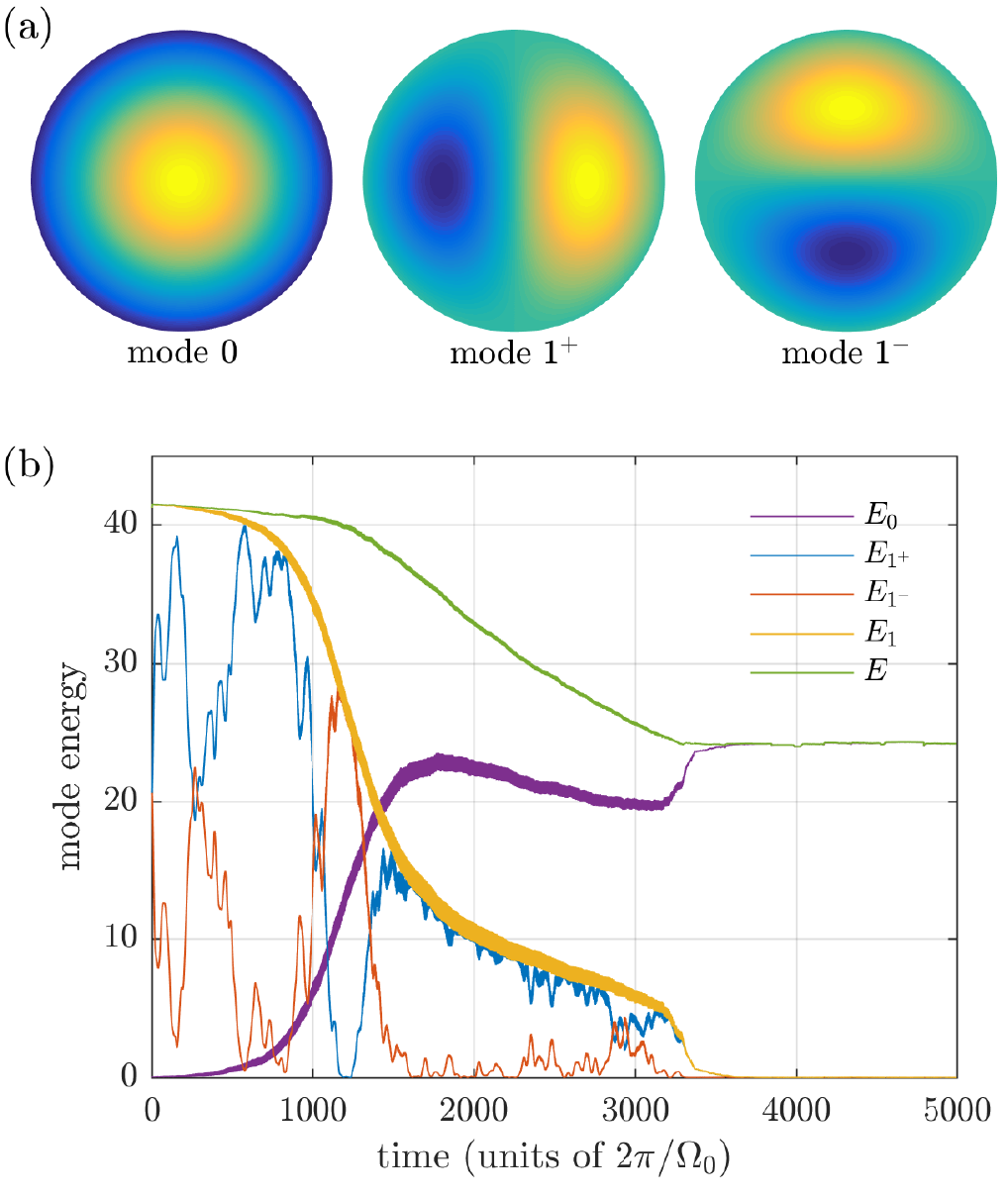}
    \caption{\textbf{(a)} The shape of the fundamental mode of a circular membrane (mode 0) along with the two degenerate first excited modes (modes $1^+$ and $1^-$). \textbf{(b)} Evolution of mode energies in a graphene membrane at 300 K. Here, $E_1=E_{1^+}+E_{1^-}$ and $E=E_0+E_1$.}
    \label{fig:modes}
\end{figure}

In this section, we introduce a 2D model where the mode functions $\phi_n({\bf x})\propto J_n(r\xi_{n,k})e^{\pm in\vartheta}$ are those of a vibrating circular membrane. The particle position ${\bf x}=(r,\vartheta)$, $r<1$, $\xi_{n,k}$ is the normalized $k$:th zero of $J_n$, and the $q_n(t)$ are the corresponding mode amplitudes. In particular, we focus on the three lowest-lying eigenmodes of the membrane [see Fig.~\ref{fig:modes}~(a)], two of which are degenerate. The three modes have frequencies $\Omega_0=\xi_{0,1}=1$ and $\Omega_{1^+}=\Omega_{1^-}=\xi_{1,1}=1.5933$. This model is described by the equations of motion
\begin{gather}
\ddot{q}_n+\Omega_n^2 q_n-\sum_{l=0,1^-,1^+} \Omega_n\Omega_l\phi_n({\bf x})\phi_l({\bf x})q_l=0,\label{eq:qem2}\\
\ddot{{\bf x}}+\Gamma\dot{{\bf x}}-\frac{1}{\mu}\sum_{j,l}\Omega_j\Omega_l q_jq_l\phi_j({\bf x})\nabla\phi_l({\bf x})=\Gamma\sqrt{2D}\xi(t).\label{eq:xem2}
\end{gather}
Here, $\xi(t)$ is  a Gaussian white noise source: $\left<\xi(t)\cdot \xi(t')\right>=\delta(t-t')$. Together with the damping coefficent $\Gamma$, the diffusion constant $D$ sets the equivalent bath temperature via the fluctuation-dissipation theorem; $T_{\rm eq.}= \mu\Gamma D/2$. 

In contrast to the case discussed in section~\ref{sec:gen}, the nodes and antinodes of different membrane modes can coincide in space, which allows for a simpler stability analysis. In the absence of noise, one finds trivially stable trapped states where the particle is located at an antinode of either mode 0 or a linear combination of modes $1^{\pm}$. Consequently, the system tends to rapidly relax to either of these states, and once the particle is trapped in a fixed position, no further relaxation takes place. Adding noise to the system, however, leads to quite different behaviors. 

The system~\eqref{eq:qem2}-\eqref{eq:xem2} bears a strong resemblance to the previously studied case~\cite{Rhen_2016} of the dynamics of a graphene drum resonator with adsorbed particles (see Appendix A). As long as the dynamics is restricted to low-lying modes where the mode frequencies remain of the same order of magnitude, the system (\ref{eq:qem2})-(\ref{eq:xem2}) provides a qualitatively accurate description of such a graphene resonator. The interest in this type of physical system, adsorbates on a vibrating drum, stems from recent years' advances in the fabrication of carbon nanomechanical resonators, which has enabled huge strides in the field of nanomechanical mass sensing~\cite{Eomreview}. The effect of fluctuations in the position of adsorbates have been detected via phase noise~\cite{Yang_2011}, but there is also numerical evidence that dissipation, as measured via ringdown~\cite{BUGuys,Rhen_2016}, is highly sensitive to particle diffusion.

\subsection{Stepwise relaxation}
In the presence of finite noise, the stepwise relaxation associated with trapping, release and retrapping occurs, as described in section \ref{sec:gen}. For numerical integration of the stochastic equations we use a second order algorithm~\cite{Mannella_1989,Mannella_2000}. Throughout this section, scaled numerical values corresponding to those of a typical micron-sized graphene drum resonator are used (see, for instance, Ref.~\onlinecite{Eriksson_2013} or Ref.~\onlinecite{Rhen_2016}). We focus on the dynamics of the system initialized in some superposition of the two degenerate modes $1^+$ and $1^-$. 

The relaxation to equilibrium in the presence of noise is depicted in Fig.~\ref{fig:modes}~(b); we can identify two regions where the total vibrational energy decays slowly. At short times, the particle is trapped near the antinode of the superposition of $1^{\pm}$, while the two modes trade energy. As fluctuations overcome the trapping, energy is transferred to mode 0 and the total energy decays until the particle is retrapped, this time at the antinode of mode 0. As was the case in section \ref{sec:gen}, the drop in total energy and intermode energy transfer is associated with the particle-mediated mode coupling, that causes the amplitude of the initially frozen mode to grow. Once enough energy has been transferred to the fundamental mode for it to dominate the dynamics -- in particular, the inertial potential becomes a single-well potential -- the relaxation proceeds as described in Ref.~\onlinecite{Rhen_2016}.

In contrast to the model studied in section~\ref{sec:gen}, the particle can become trapped at any energy, and there are thus no preferred energies where the plateaus appear. 
\begin{figure}
    \centering
    \includegraphics[width=\columnwidth]{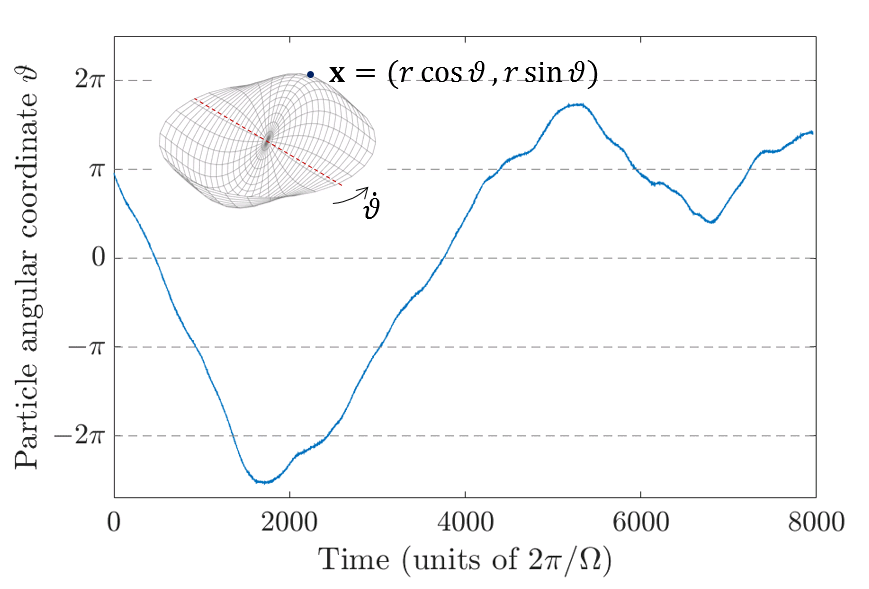}
    \caption{Noise-induced angular precession. When the first excited mode is active, and the particle is trapped at an antinode, noise induces stochastic precession of the angular coordinate $\vartheta$ of the particle position. The position of the node-line (and the antinode) follows this precession. The timescale over which the sense of direction of precession changes is typically much larger than the timescale of vibration.}
    \label{fig:precess1}
\end{figure}

\subsection{Stochastic precession}
If the amplitude of mode 0 is much lower than the amplitude of mode 1, and the particle is trapped at the antinode of mode 1, thermal fluctuations in its position causes a precession in the angular direction. This precession is manifested in the fluctuations in mode energies seen in the short time behavior in Fig.~\ref{fig:modes}~(b) for $t<1000\times 2\pi/\Omega_0$. While this membrane model shares with the one in Sec.~\ref{sec:gen} that the degenerate modes trade large amounts of energy, the stochastic fluctuations seen here contrast with coherent oscillations that previously appeared. Here, these fluctuations of relative amplitude of the two modes result in a random precession of the node/antinode of the vibration, and consequently of the angular coordinate $\vartheta$ where the particle is trapped (see Fig.~\ref{fig:precess1}). 

\begin{figure}
    \centering
    \includegraphics[width=\columnwidth]{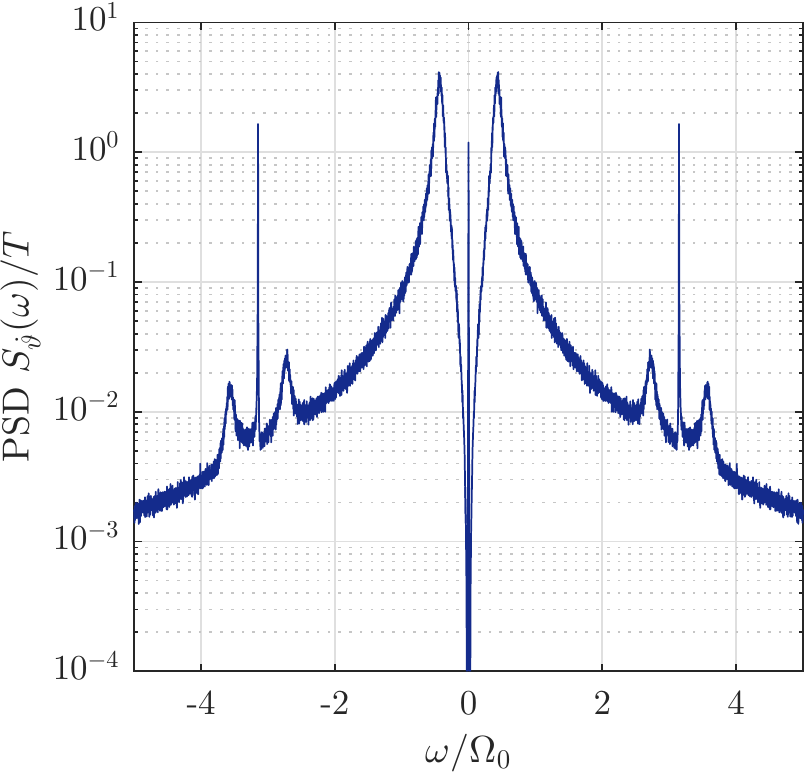}
    \caption{Normalized power spectral density (PSD) of precession velocity $S=|\omega \hat{\vartheta}(\omega)|^2$. Clearly visible are sharp peaks around $\omega=2\Omega_1$ associated with the parametrically driven nonresonant motion. The sidebands come from mixing with the irregular quasiperiodic particle motion around $\omega\approx 0.44 \Omega_0$. The low frequency part is dominated by a Lorentzian peak; see Fig.\ref{fig:precess3}~(a).}
    \label{fig:precess2}
\end{figure}

To characterize the stochastic precession, it is convenient to consider the power spectral density (PSD) for angular velocity, $S_{\dot\vartheta}=|\omega \vartheta(\omega)|^2$, obtained from the time series of $\vartheta(t)$. A temperature-normalized PSD is shown in Fig.~\ref{fig:precess2}, from simulation at $T=100$~K. Clearly visible are the narrow parametric resonant peaks at $2\Omega_{1^{\pm}}\approx 3.2 \Omega_0$, along with sidebands stemming from mixing with the particle's quasiperiodic motion in the trapping potential at $\omega\approx 0.44\Omega_0$. 

\begin{figure}
    \centering
    \includegraphics[width=\columnwidth]{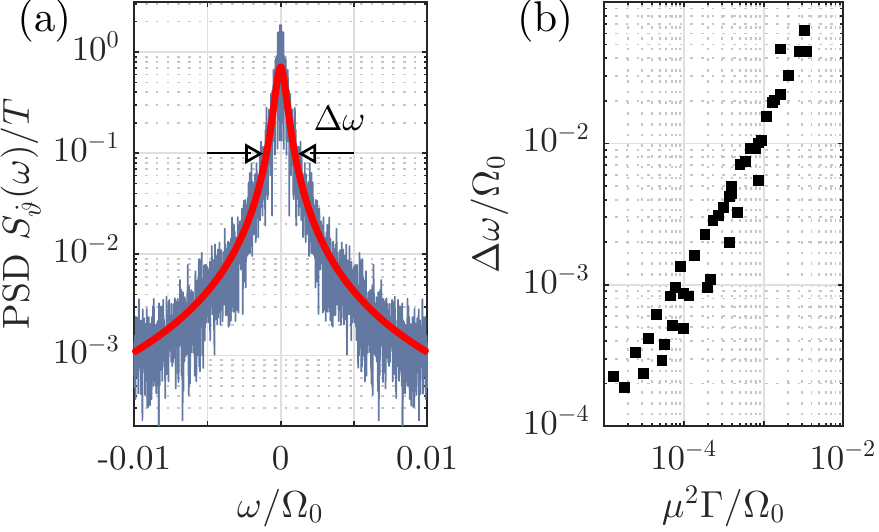}
    \caption{Characterization of low-frequency precession noise at $T_{\rm eq.}=100$~K. (a) Close up of the central frequency noise PSD in precession velocity in Fig.~\ref{fig:precess2}. The red line is a fit to a Lorentzian with a width $\Delta\omega \ll \Gamma$. (b) Scaling of width $\Delta\omega$ with dissipation and added mass. Varying $\Gamma$ and  $\mu$ reveals an approximately linear relationship between $\Delta\omega$ and $\mu^2\Gamma$. }
    \label{fig:precess3}
\end{figure}

Measurements at high frequencies are usually hard, and more interesting is thus the low-frequency noise centered around $\omega=0$. A close-up of the peak around zero frequency is shown in Fig.~\ref{fig:precess3}~(a), along with a Lorentzian fit. The width of the peak, $\Delta\omega$, gives the characteristic timescale for the low-frequency noise. Numerical simulations reveal that this width scales as $\Delta\omega \sim \mu^2\Gamma$ [see Fig.~\ref{fig:precess3}~(b)] for fixed temperature (recall $T_{\rm eq.}= \mu\Gamma D/2$). The result is a timescale which is much longer than one set by the particle damping rate $\Gamma$. Indeed, using dimensionfull units, for a particle of mass $m$ on a vibrating membrane of radius $R$ and mass $M$, this corresponds, in the low-frequency domain, to the mixing angle performing a random walk with RMS-angular velocity $\langle\dot{\vartheta}^2\rangle^{1/2}\sim R^{-1}(mk_{\rm B}T/M^2)^{1/2}$. This stochastic mode precession will cause noise in the measurement of the mode amplitude if one uses a split back gate that only couples to one of the modes. Such geometries were recently used for controlled actuation and readout of individual modes and mode shapes in, for instance, graphene resonators~\cite{Delft, Cornell}.


\section{Conclusions}
Inspired by the coupling between diffusive degrees of freedom and oscillator systems, we have studied the free evolution and relaxation dynamics in two systems belonging to the more generic class of systems, given by the Hamiltonian \eqref{eq:h0}-\eqref{eq:hi}. In particular, we studied the case when degenerate modes are present. Using a minimal 1D prototype system, we demonstrated a characteristic, stepwise relaxation. These steps are associated with trapping, release and subsequent retrapping of the particle(s) in the interaction potential. 

As a concrete example, we then demonstrated that the same characteristic behavior can be seen in 2D membrane resonators with adsorbed particles. However, in the latter case, thermal noise is necessary to facilitate the release from the trapping. Finally, we demonstrated that for the membrane model, when the particle is trapped by a degenerate mode, thermal noise induces a slow-in-time stochastic precession of the mode-mixing angle.

The results in this paper highlights how the approach to equilibrium can have a highly nontrivial behavior, departing from the usual exponential relaxation, also in very simple physical model systems.  
\begin{acknowledgments}
We acknowledge financial support from the Swedish Research Council (VR) and Knut and Alice Wallenberg foundation (KAW).
\end{acknowledgments}

\appendix

\section{Mode coupling matrix for a nanomechanical drum resonator\label{app:matrix}}
In its simplest incarnation, a circular mechanical resonator made from graphene or some other 2D material, can be modeled as membrane suspended in the $xy$-plane with a tension $\sigma$. Denoting the deflection in the $z$-direction by $w(x,t)$, the equation of motion for such a membrane is
\begin{equation}
(\rho+\delta\rho)\partial_t^2{w}-\sigma\nabla^2 w=0.
\end{equation}
Here, $\rho$ and $\delta \rho$ are the intrisic membrane sheet density and density fluctuations due to the particles: $\delta \rho=m\sum_k \delta(x-x_k)$. It is here assumed that the adsorbed particles are much lighter than the membrane itself, and that their motional timescale satisfies $(m/M)\dot{x}_k\ll \partial_t w$. For sufficiently light particles, this condition can always be fulfilled. The corresponding equation for the particle motion reads
\begin{equation}
m\ddot {x}_k+m\Gamma\dot{x}_k+m\ddot{w}(x_k)\nabla w(x_k)=f_{\rm ext}.
\end{equation}

Expanding the membrane motion into eigenmodes $w(x,t)=\sum_n q_n(t)\phi_n(x)$ yields
\begin{gather}
\ddot{q}_n+\Omega_n^2q_n+\epsilon\sum_{m,k} \ddot{q}_m \phi_m(x_k)\phi_n(x_k)=0,\\
\ddot{x}_k+\Gamma\dot{x}_k+\sum_{m,n}\ddot{q}_nq_m \phi_m(x_k)\nabla\phi_n(x_k)=f_{\rm ext}/m.
\end{gather}
While formally correct, the system of equations is cumbersome to treat due to the time-varying quantity $\delta\rho\ddot{q}_n$. However, we are typically concerned with small perturbations to harmonic motion, in which case we can, to lowest order in $\epsilon=m/M$, approximate $\ddot{q}_n\approx -\Omega_n^2 q_n$. 
This approximation, together with a change of scale $q_n\rightarrow q_n/\Omega_n$ allows the equations of motion to be written as
\begin{gather}
\ddot{q}_n+\Omega_n^2q_n-\sum_{m,k} g_{mn}q_m \phi_m(x_k)\phi_n(x_k)=0,\label{eq:appq}\\
\ddot{x}_k+\Gamma{x}_k-\frac{1}{\epsilon}\sum_{m,n}\frac{g_{mn}}{\Omega_n^2}{q}_nq_m \phi_m(x_k)\nabla\phi_n(x_k)=f_{\rm ext}/m.\label{eq:appx}
\end{gather}
Here, $g_{mn}=\epsilon\Omega_n\Omega_m$. These equations are of the same form as those following from the Hamiltonian~\eqref{eq:h0}-\eqref{eq:hi}, and for the fully degenerate case $\omega_n\equiv \omega$, Eqs.~\eqref{eq:appq}-\eqref{eq:appx} are indeed the Hamiltonian equations of motion. For the general case, however, the inertial force acting on the particles in Eq.~\eqref{eq:appx} does not follow from the simple Hamiltonian. This is problematic, as energy conservation is no longer guaranteed in the absence of damping and noise. 

As we are here interested in a qualitative picture of the dynamic, we limit the membrane study to the 3 lowest modes, out of which two are degenerate. In this case, the ratio of the two frequencies is still of order unity ($\Omega_0/\Omega_1^{\pm} \approx 0.63$), and replacing $g_{mn}/\Omega_n^2 \rightarrow g_{mn}$ thus somewhat underestimates the inertial force associated with acceleration from the fundamental mode. 

\end{document}